\documentclass{caosp309}

\usepackage{graphicx}

\usepackage{natbib}
\articleNo{xx}
\pubyear{2024}
\volume{xx}
\volnumber{x}
\firstpage{1}
\received{Nov 15, 2024}
\accepted{Dec 1, 2024}

\def\BibTeX{{\rm B\kern-.05em{\sc i\kern-.025em b}\kern-.08em
             T\kern-.1667em\lower.7ex\hbox{E}\kern-.125emX}}

\begin{document}

%
\hauthor{Zs.\,K\H{o}v\'ari}

\title{Magnetic activity in close binary systems}


%
%
\author{
        Zs.\,K\H{o}v\'ari\inst{1,2}\orcid{0000-0001-5160-307X}
       }

%
\institute{
           Konkoly Observatory, HUN-REN Research Centre for Astronomy and Earth Sciences, H-1121 Budapest, Konkoly Thege \'ut 15-17., Hungary \email{kovari@konkoly.hu}
           \and
           HUN-REN CSFK, MTA Centre of Excellence, H-1121 Budapest, Konkoly Thege \'ut 15-17., Hungary
          }

\date{Nov 15, 2024}

\maketitle

\begin{abstract}
Tidal forces in close binary systems have diverse impacts on magnetic activity. The synchronicity characteristic of close systems counteracts magnetic braking, thereby sustaining rapid rotation—a key factor in increased levels of magnetic activity. Tidal effects can also work against the slowing down of rotation during stellar evolution, when the star inflates into a red giant. A notable manifestation of the effect of binarity on activity in such systems is the appearance of active longitudes, which are thought to arise from the excitation of non-axisymmetric dynamo modes. Through some recent examples, the dinamo operation in RS\,CVn and BY\,Dra type systems is briefly reviewed in terms of spot cycles, active longitudes, flare activity, and differential rotation.

\end{abstract}

%
\section{Introduction}
\label{intr}
Indicators of magnetic activity known on the Sun, such as dark photospheric spots, bright chromospheric plages, flares and coronal X-ray and radio emission, are common in late-type stars with convective envelopes. Detached close binary systems with such an active late-type component (e.g., RS\,CVn or BY\,Dra type systems) provide a unique environment where magnetic fields, tidal forces, and stellar dynamics interact in highly complex and extreme ways. The proximity of the stars can amplify the effects of magnetism, driving phenomena such as intense starspot activity, frequent flaring, and strong chromospheric and coronal emissions. These magnetic processes not only affect the observable characteristics of the stars but also play a critical role in shaping the long-term evolution of the binary system itself.

One of the key elements that makes the study of magnetic activity in close binaries so compelling is the multifaceted role of tidal interactions. Tidal synchronization can enhance magnetic activity by speeding up rotation, which in turn powers the dynamo processes that generate strong magnetic fields. The long-period group of RS\,CVn binaries typically consists of such systems with evolved but still rapidly rotating active components. Moreover, tidal forces distort the stars' shapes, which introduces non-axisymmetry, yielding variations in the internal plasma flows and surface conditions, ultimately altering the magnetic dynamo \citep[see, e.g.,][]{1997A&A...321..151M,2002MNRAS.334..925S,2002A&A...396..885M}. This affects observable signs of magnetic activity, such as the distribution of starspots \citep[e.g.,][]{2024MNRAS.529.4442S}, but may also contribute to the modulation of long-term magnetic behavior. However, tidal interactions also play a central role in orbital evolution, as the angular momentum exchange driven by tides and magnetic braking can lead to significant changes in the orbital period, eventually causing the stars to spiral closer together or even merge.

The combination of magnetic activity and tidal interactions creates a dynamic and evolving system, where changes in one aspect (such as rotation or orbital configuration) can ripple through and affect the magnetic, radiative, and mechanical properties of the stars. Therefore, by studying these systems, we can gain deeper insights into stellar magnetic dynamos, the role of magnetic activity in stellar evolution, and the complex interplay between stellar interiors, surfaces, and orbital dynamics. But understanding these processes in close binaries has broader implications, such as for the study of solar activity or even the evolution of exoplanetary systems around active stars.

In the recent past, high-resolution spectrographs, and space-based observatories, has significantly advanced the study of active close binaries. The surface of the spotted components can be studied in great detail with the Doppler imaging technique, which provides sufficiently robust and reliable results with adequate time (rotational phase) coverage, in contrast to photometric spot models, where the mapping of the surface of the star is only possible to a very limited extent \citep[e.g.][]{1997A&A...323..801K,2021AJ....162..123L}.

In the following review, we go through the most prominent observable characteristics of magnetic dynamos operating in detached close binary systems with the help of selected examples from recent observational studies. However, due to space limitations, we cannot deal with the semi-detached or contact binaries (Algols, W\,UMa systems) in this review (but see a former review on the topic by \citealt{1993ASSL..177...51G}, or a more recent statistical study by \citealt{2019PASJ...71...21K}), nor can we cover the theoretical foundations of the topic. Readers interested in the latter can be recommended, e.g., the series of papers by \citet{1991A&A...251..183S} and \citet{2000AN....321..175H,2003A&A...405..291H,2003A&A...405..303H}.

\section{Observables of magnetic activity in close binaries}

\subsection{Starspots}
Starspots are the most prominent manifestations of magnetic fields on the surface of stars. Their role is crucial for understanding stellar magnetic activity. By studying them we gain insights into the underlying stellar dynamos. But in binary systems, starspots provide clues also about the interactions between the stars, for example in relation to rotational synchronization by tidal forces. Moreover, from the observation of the starspots, it is even possible to infer the interbinary magnetic interactions, at least to some extent, which can affect orbital dynamics as well. However, there is no doubt that monitoring starspot activity over time helps in understanding magnetic cycles and ultimately the mechanisms that generate and sustain magnetic fields.

Starspots can be observed using different methods of which we highlight the three most important here. \emph{Photometric spot modeling} of light curves \citep[see, e.g., the summary by][]{2013IAUS..294..257S} reveals spots as they rotate in and out of view, but the time series (i.e. 1-dimensional information) is not sufficient for spatial resolution. High-resolution spectroscopy, on the other hand, offers a more detailed approach through the \emph{Doppler imaging} technique \citep[e.g.,][see also their references]{2008A&A...488..781C,2012A&A...548A..95C}, where distortions in spectral lines caused by the star’s rotation help create surface maps showing starspot locations. In more advanced observations, long-baseline optical/near-infrared \emph{interferometry} allows for direct imaging of a star's surface, resolving starspots as dark regions, though this technique relies on highly specialized instrumentation and has strong constraints regarding the target itself. For a comparison of simultaneous visualizations of the spotted surface of the giant component in the RS\,CVn system $\sigma$\,Gem made with these three different imaging techniques, we cite the work of \citet{2017ApJ...849..120R}.

\subsection{Active longitudes}

The formation spot concentrations along a certain longitude range (but sometimes even at high latitudes around the rotational poles), with lifetimes of the order of months to years, that is active or preferred longitudes, are commonly observed phenomena in active close binaries \citep{1998A&A...336L..25B}. In synchronized systems these activity centres can form and also bind to certain orbital phases \citep[e.g.,][]{2006Ap&SS.304..145O}. Active longitudes, however, require breaking the axial symmetry. This is consistent with the expectation of non-axisymmetric dynamo modes due to symmetry-breaking tidal forces \citep{1997A&A...321..151M}. Another condition for the excitation of stable, non-axisymmetric fields is that the differential rotation is not too strong \citep{1995A&A...294..155M}. We note that this has been verified for synchronously rotating components in close binaries in many cases \citep{2017AN....338..903K}, which is not surprising because tidal coupling is expected to diminish differential rotation.

Certainly, the long-term monitoring of these structures provide insight into the large-scale structure and evolution of the stellar magnetic fields. The location of the active longitudes in the reference frame of the orbit was investigated by \citet{2006Ap&SS.304..145O} for 12 close binaries. According to the study, active longitudes on giants in RS\,CVn-type systems can typically be linked to substellar positions, however, in the case of active subgiant components, the picture is not so clear, rather mixed. At the same time, for main-sequence stars of BY\,Dra type systems, the spots are mostly concentrated in quadrature positions. However, other studies led to somewhat contradictory results \citep[cf.][]{1998A&A...336L..25B,2007AcA....57..347S,2024MNRAS.529.4442S}.

\subsection{Activity cycles}

The solar activity cycle is recorded for hundreds of years, and this database can be extended to orders of magnitude longer timescales using various proxies recorded on Earth that reflect changes in solar activity. However, the long-term data collected for active close binaries cannot yet be compared to that of the Sun, either in terms of the time base or the regularity. Nevertheless, from the long-term brightness fluctuations observed in RS\,CVn and BY\,Dra type stars were explained long ago by the fact that the active components in such binaries rotate relatively rapidly as a result of tidal synchronization, which ultimately manifests itself in strong chromospheric activity. In other words, from the point of view of activity cycles, binarity has a role mostly in creating/maintaining rapid rotation. That is, the difference is not so much between an active single star and a star similar to it but in a binary system. In terms of dynamo operation, rotation, differential rotation and the evolutionary stage play a more important role \citep[cf.][]{1996ApJ...460..848B,2017LRSP...14....4B,2023SSRv..219...58K}.

In a recent statistical study by \citet{2022MNRAS.512.4835M}, the authors examined 120 RS\,CVn-type (but including some BY\,Dra-type) stars from the Southern Hemisphere, and for 91 systems they were able to detect activity cycles. In fact, however, the activity cycles determined so far for active stars (again: not only strictly speaking for members in close binary systems) usually seem to be time-varying quasi-cycles, i.e. instead of precisely defined periods, only approximate averages can be determined. Moreover, in many cases multiple cycles were detected—just like on the Sun. As a reference here are some studies that contain also close binaries in a significant part: \citet{2009A&A...495..287B,2009A&A...501..703O,2013AN....334..625O,2021PASA...38...27O}.

\subsection{Active coronae and flares}
In the most accepted solar flare model \citep[e.g.,][]{1976SoPh...50...85K}, the release of magnetic energy is caused by the reconnection of magnetic field lines. In addition, however, other possible mechanisms are known to induce reconnection in either the solar, or stellar magnetospheres, see e.g., in \citet{1979SoPh...64..303H,2001SoPh..203..321C,2010hssr.book..159F,2014masu.book.....P}, etc.
However, in close binaries, the binary nature itself plays the decisive role in the formation of flares in many cases. Radio and X-ray observations of close binaries suggest that in several such systems the extent of magnetic structures is comparable to the separation between the companion stars \citep[e.g.,][etc.]{1985ASSL..116..275L,1996ApJ...473..470S,2010Natur.463..207P,2020IAUS..354..418H}. This indicates that the magnetic activity is not only bound to one or the other component, but is actually of interbinary origin. That is, we see the direct result of the magnetic interaction between the components. One such characteristic interaction is when the corotating giant magnetic loops present on both stars are large enough to reach each other and become entangled, and therefore the flux tubes can be temporarily bridged \citep{1980ApJ...239..911S}.

The role of interbinary magnetic fields in flare activity is still not fully understood. However, it is very likely that these interacting fields are not only responsible for the total quiescent X-ray light, but may also be the source of large, long-lasting energetic X-ray flares, as suggested by \citet{2022ApJ...934...20S}. All of this seems to be supported by the recent finding by \citep{2021A&A...647A..62O}, according to which 
it is reasonable to suspect that a large fraction of red giants showing enhanced flare activity found in the \emph{Kepler} field do belong to close binary systems.

We note however, that it is not absolutely necessary to assume an interbinary magnetic entanglement since rapid rotation maintained by synchronization alone can increase the level of flare activity of such red giant components. Moreover, other types of interaction processes are also possible; e.g. \citet{1988A&A...205..167V} long ago drew attention to the fundamental role of the proximity of the companion star in the formation of filament support and flares in close binaries.


\section{V471\,Tau: a poster child for active close binaries}

We illustrate the effect of close binarity on the magnetic activity through the example of V471\,Tau, an eclipsing binary system of an active K2 red dwarf star and a white dwarf, i.e., a post-common envelope binary, also, a pre-cataclysmic system. V471\,Tau is an actual astrophysical laboratory for studying many aspects of stellar evolution \citep[what's more, the system is probably triple, see][]{2022MNRAS.517.5358K}. To reveal the main characteristics of the K star's magnetic activity, in our recent study \citep{2021A&A...650A.158K} we used space photometry from \emph{TESS}, high-resolution spectroscopy from the CFHT-ESPaDOnS data archive\footnote{https://www.cadc.hia.nrc.gc.ca/AdvancedSearch/}, and X-ray observations from various space instruments. Note, that in such post-common envelope binary systems X-ray emission can originate from either the white dwarf or the corona of the active K star.

\begin{figure}[t]
    \centering
    \includegraphics[width=0.7\textwidth]{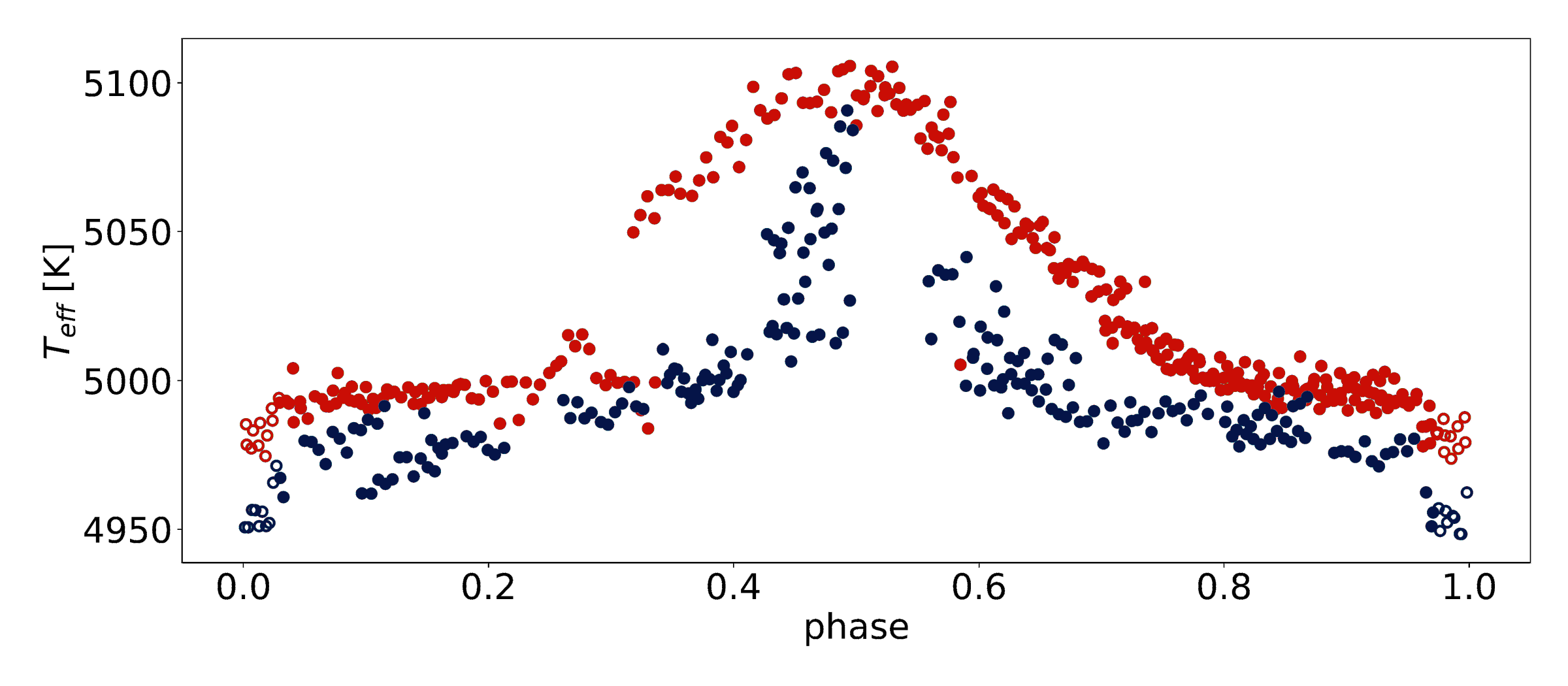}
    \vspace{-0.35cm}
    \caption{Variation of the surface temperature of the K2V component of V471\,Tau along the orbital phase due to the irradiation effect. Temperature values were derived individually for each observed spectrum (red dots for the 2005 season, black dots for 2014/15). Open circles around zero phase correspond to those unaffected spectra that were observed when the white dwarf was eclipsed.
    }
    \label{fig1}
\end{figure}

\begin{figure}[thb]
    \centering
    \includegraphics[width=0.85\textwidth]{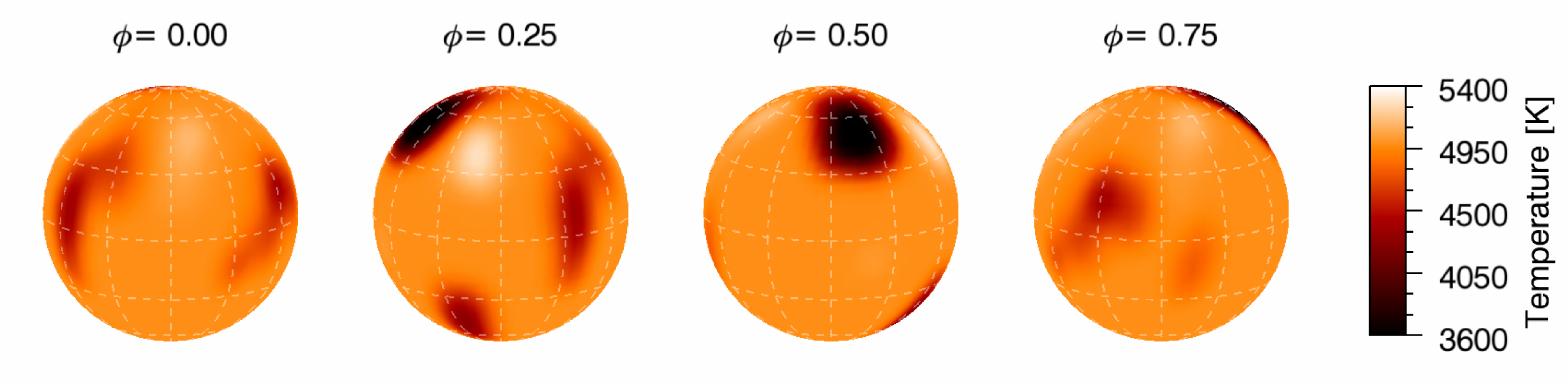}
    \includegraphics[width=0.85\textwidth]{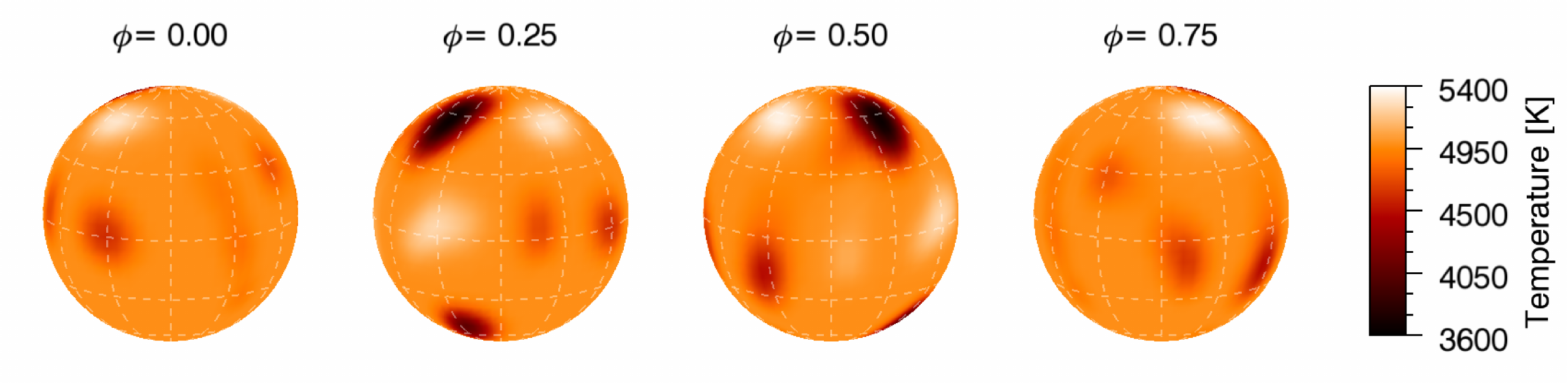}
    \vspace{-0.35cm}
    \caption{Doppler image of the K2V component of V471\,Tau from 2005 data (top panel) and from 2014/15 data (bottom panel). Both surface temperature maps are plotted in quarter rotation phases;  $\phi$=0.0 corresponds to the phase when the white dwarf is eclipsed, $\phi$=0.5 is the phase facing the white dwarf. Note the stable spot configuration despite a decade of time difference between the images.
    }
    \label{fig2}
\end{figure}

The rotation of the K2 dwarf is locked tidally to the orbital period of the eclipsing system of $\approx$0.52\,d. One of the direct consequences of tidal locking is that the hemisphere of the K-star facing the white dwarf is constantly exposed to the hot star's radiation, which heats it up. On the basis of spectral synthesis we found that the surface temperature of the K star indeed varies along the orbital phase: an average temperature rise of $\approx$100\,K around $\phi$=0.5 phase (so when the white dwarf is in front) indicates the irradiance effect in both the 2005 and 2014/15 observing seasons; see Fig.\,\ref{fig1}. Meanwhile, the hemisphere of the K2 star facing the hot component is more covered with cool spots; see our Fig.\,\ref{fig2} for the results of Doppler imaging for both seasons. Moreover, despite the ten-year difference between the Doppler images shown in the figure, the surface spot distribution of the star K2 is remarkably stable. It is particularly noteworthy that the phase of the most dominant spot essentially coincides with the phase facing the white dwarf. This permanent cool spot, or say active longitude, may be coupled to the orbital phase \citep[cf.][]{1991A&A...251..183S}.


\begin{figure}[hb!]
    \centering
    \includegraphics[width=0.75\textwidth]{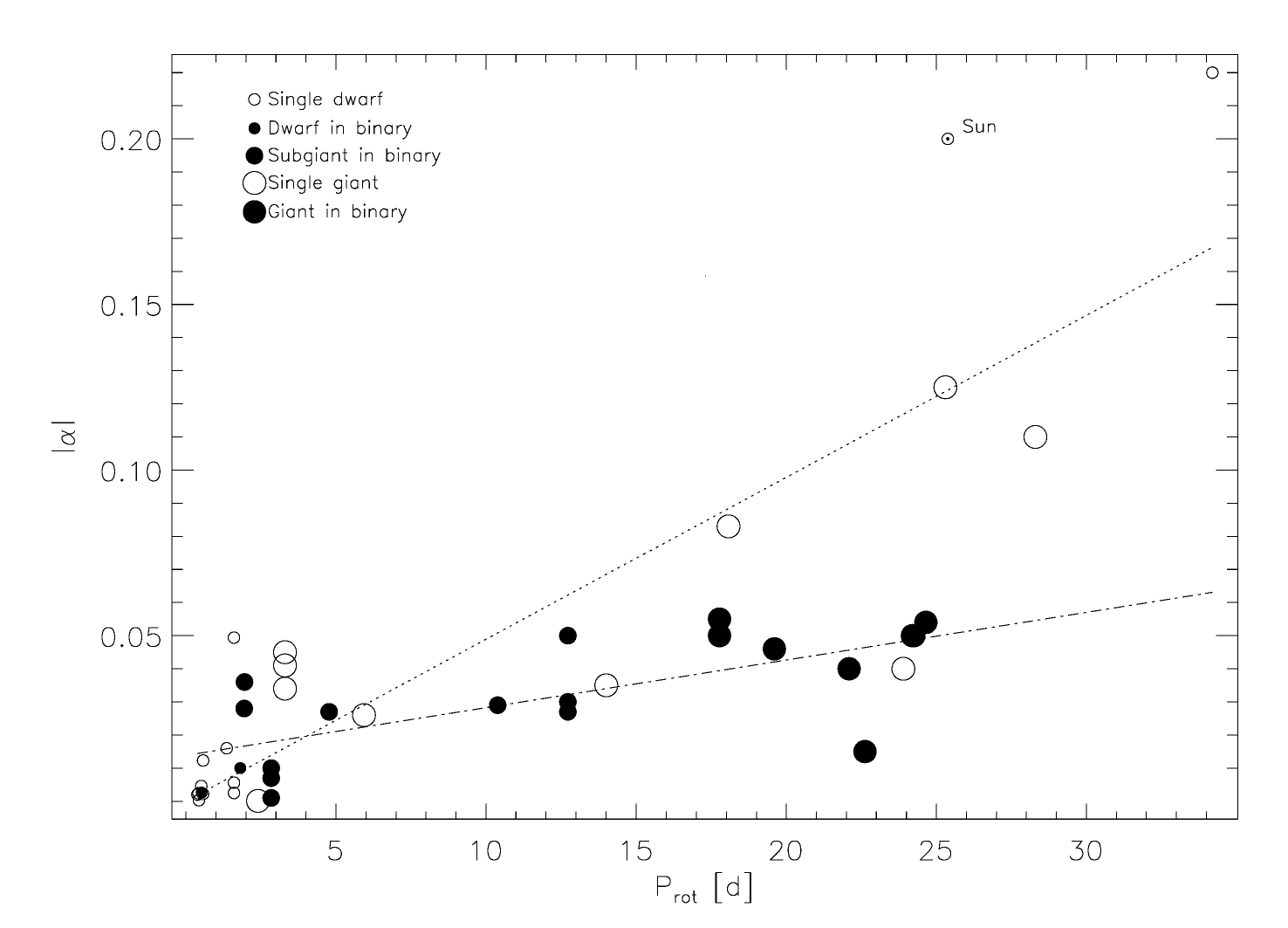}
    \vspace{-0.35cm}
    \caption{Surface shear parameter ($\alpha$) vs. rotation period ($P_{\rm rot}$) graph for active spotted stars \citep[updated from][]{2017AN....338..903K}. Open circles fitted with a dashed straight line correspond to single active stars, filled black circles fitted with a dash-dotted straight line represent active stars in close binary systems. Symbol size increases from dwarfs to subgiants to giants.
    }
    \label{fig3}
\end{figure}

Our thorough Doppler study in \citet{2021A&A...650A.158K} also provided the opportunity to follow the changes in the spot distribution within a short period of time (i.e., within a few days or a few weeks). The cross-correlation technique presented therein is one of the best known and most established ways to measure the differential rotation of the stellar surface \citep{1997MNRAS.291....1D}. Briefly, surface differential rotation can be measured by cross-correlation of the consecutive Doppler images along the astrographic longitude. On the resulting cross-correlation function map, the strongest correlation pattern gives the most probable latitudinal velocity distribution. The pattern can be fitted with a quadratic rotation law in the usual form known from the solar case:
\begin{equation}\label{eq:drlaw}
    \Omega(\beta)=\Omega_{\mathrm{eq}}(1-\alpha_{\mathrm{DR}}\sin^2\beta),
\end{equation}
where $\Omega(\beta)$ is the angular velocity at $\beta$ latitude, $\Omega_{\mathrm{eq}}$ is the angular velocity of the equator, and $\alpha_{\mathrm{DR}}$ is the dimensionless surface shear parameter. An improved version of the technique \citep[dubbed \texttt{ACCORD}, see e.g.,][]{2004A&A...417.1047K,2012A&A...539A..50K,2015A&A...573A..98K}, which uses the average of cross-correlations, is even more robust, i.e. less sensitive to rapid stochastic changes in the arrangement of spots.

For the K2 star of V471\,Tau, the surface shear parameter was found to be $\alpha_{\mathrm{DR}}=0.0026\pm0.0006$, which indicates an almost solid body rotation. This finding confirms the assumption that in the case of such a close binary, tidal locking inhibits the differential rotation. Further evidence of such tidal confinement can be seen in Fig.\,\ref{fig3}, which compares the shear parameters for various single and close binary stars with different rotation periods. The plot is an updated version of the surface shear parameter vs. rotation period graph ($|\alpha|$--$P_{\rm rot}$ graph) from \citet{2017AN....338..903K}. The dotted and dash-dotted lines denote the linear fits for single stars and stars in close binary systems, respectively, with slopes of $|\alpha|\propto 0.0049P_{\rm rot}$[d] and $|\alpha|\propto0.0014P_{\rm rot}$[d]. The diagram clearly indicates bimodality, i.e., that close binarity does play a role in inhibiting the differential rotation of the convective envelope.

\begin{figure}[t]
    \centering
    \includegraphics[width=0.75\textwidth]{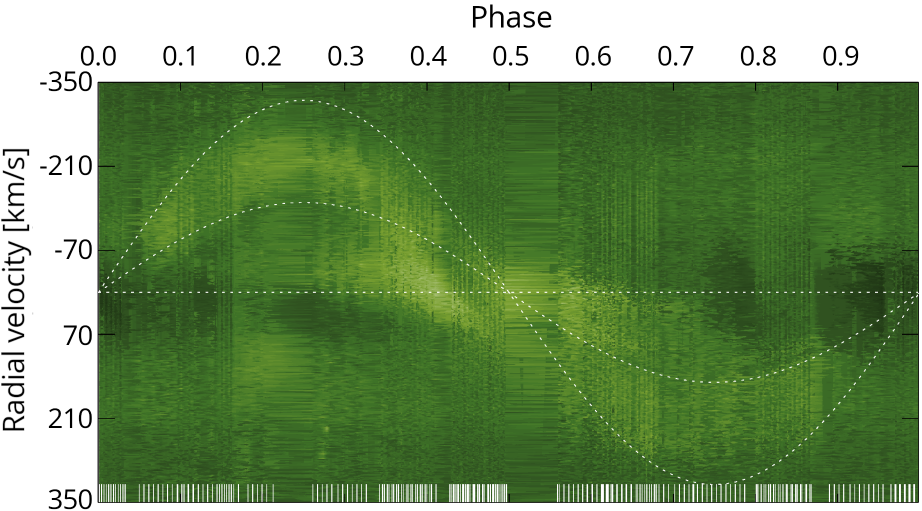}
    \vspace{-0.00cm}
    \caption{Dynamic H$\alpha$ spectrum of V471\,Tau from the 2014/15 data. The individual 1-D H$\alpha$ spectra along the orbital phase are plotted in the rest frame of the K2 star. Dark green corresponds to deep absorption, while light yellow indicates strong emission. The large amplitude ($\approx$320\,km\,s$^{-1}$) sine curve marks the radial velocity of the white dwarf, the smaller amplitude ($\approx$150\,km\,s$^{-1}$) sinusoid is the radial velocity of the centre of mass of the binary system. Tick marks at the bottom indicate the phases of the observations.
    }
    \label{fig4}
\end{figure}

In connection with V471\,Tau, we mention one more activity characteristic that raises the importance of close binarity: the inter-binary H$\alpha$ emission in 2014/15. Based on long-term X-ray observations, the star was close to its maximum activity during this period \citep[][see their Fig.\,13]{2021A&A...650A.158K}. The H$\alpha$ behaviour in Fig.\,\ref{fig4} shows absorption centered around 0.0 phase and the emission between 0.35-0.65 phase, this latter around the phase of the most prominent cool spot (cf. Fig.\,\ref{fig2}). Being within the radial velocity range of the K2 dwarf, these features are localized on the surface of the red dwarf. However, there is a very different emitting phenomenon located between the centre of mass of the system and the white dwarf. This emission is stronger in the first half of the orbital phase, but still visible during the second half. This suggests the presence of a clump of emitting plasma corotating with the system in the vicinity of the inner Lagrangian (L1) point, as already reported by \citet[][see their Fig. 3]{1991ApJ...378L..25Y}, who also found external H$\alpha$ emission component, originated from the inter-binary space beyond the centre of mass of the system, expanding towards the white dwarf, very similar to our result plotted in Fig.\,\ref{fig4}.

It is quite possible that the extra H$\alpha$ emission from the inter-binary space might have something to do with the enhanced X-ray luminosity in 2015, i.e., around activity maximum. We suspect that during the magnetic cycle, around activity maximum more and/or more extended magnetic loops develop high above the surface of the K2 star, reaching one stellar radius \citep{1986ESASP.263..197G}, and this way more cool material enters the upper atmosphere in trapped clumps, similar to solar prominences \citep[cf.][]{1989MNRAS.238..657C}. Normally, such a clump of cool material appears in absorption in H$\alpha$, exactly what we actually see on the `far side' of the K2 star around zero phase. However, it can get into emission when heated due to the UV radiation from the white dwarf. The physical process behind heating proposed first by \citet{1988ApJ...334..397Y} involves fluorescence-induced H$\alpha$ emission \citep[see also][]{1991AJ....102.2079B}. This scenario could explain the H$\alpha$ line variability around activity maximum (Fig.\,\ref{fig4}) when extended loops with cool material trapped in them are present.

Finally, we mention here a special group of active binaries: those close binaries in which both components are magnetically active, so that their magnetospheres can interact directly with each other \citep{1997eaun.book..151O}. In such cases, the rotation and orbital motion of the stars can lead to a slow overall winding of the coupled magnetic fields, allowing a gradual accumulation of magnetic energy that can eventually be released as an interbinary flare \citep[][]{2021ApJ...923...13C}. Note that something similar happened in UX\,Ari (G5V+K0IV), when the corotating, giant magnetic loops present on both stars interacted and flux tubes temporarily connected the components \citep{1980ApJ...239..911S}. Such connecting flux loops can also pave the way for mass transfer between the two stellar components \citep[see, e.g.][]{2005A&A...433.1055F}, which can even be observed in the form of orbital period changes \citep[cf.,][]{1979ApJ...230..815D}. Another interesting target is DQ\,Tau, in which the orbit of the two $\sim$0.65$M_{\odot}$ mass pre-main sequence components is highly eccentric, therefore at periastron encounters the magnetospheres of the components collide and trigger each other, resulting in magnetic reconnection and so, as a result, recurring flares can be observed in these orbital phases \citep{2010A&A...521A..32S}.

\section{Active red giants in long period RS\,CVn systems}

In long period RS\,CVn systems with an active cool giant component, tidal synchronization helps maintain rapid rotation, even though blowing up to a giant star would just slow down the rotation. This process has a fundamental influence on dynamo operation, which is illustrated by some examples of great importance.

One notable target is $\sigma$\,Gem, a close binary with an active red giant primary that rotates synchronously with the orbital period ($P_{\rm rot}\approx P_{\rm orb}\approx$20\,d). What makes the star really interesting is that its differential rotation was found to be antisolar, i.e. the equatorial band rotates the slowest \citep[$\alpha$=$-$0.04$\pm$0.01][]{}. We note that so far only a few stars have been confirmed to exhibit antisolar differential rotation and there is good reason to assume that these are mostly subgiants or giants \citep[cf.][see their Table 1]{2017AN....338..903K}. \citet{2015A&A...573A..98K} also reported a significant poleward flow of $\sim$0.3km\,s$^{-1}$ for $\sigma$\,Gem, which was measured using spot tracking. According to \citet{2004AN....325..496K} antisolar differential rotation may result from a fast meridional flow, if some driving mechanism is able to maintain it. Such driver candidates are large-scale thermal inhomogeneities (i.e., large starspots) and/or tidal effects in close binary systems. $\sigma$\,Gem meets both criteria, so the measurements and the theoretical basis verify each other.

A common feature of long period RS\,CVn giants
is that large starspots develop on their surface from time to time, which can persist for several months and can cause modulation with an amplitude of up to $\sim$0.3-0.8\,mag in the $V$ band. However, it is not common for these activity centres to be bound to the orbit. The most recent long-baseline optical and infrared interferometric starspot observations of $\zeta$\,And have revealed beyond any doubt that the spots can also appear at high latitudes, even covering the rotation poles. In addition, significant asymmetries can occur between the hemispheres regarding the appearance of spots \citep[cf.][]{2016Natur.533..217R}. This suggests highly asymmetric dynamo modes, i.e. a dynamo operation other than the Sun's, e.g., with mixed parity modes \citep{2003MNRAS.338..655B}. What is more, decades-long photometry of such overactive RS\,CVn giants showed that their long-term peak-to-peak brightness variation can be as much as one magnitude \citep{2014A&A...572A..94O}, and these long-term changes are defintely not periodic, often a mixture of aperiodic and trend-like changes. Based on all this, it can be concluded that the dynamo of such RS\,CVn giants is non-solar in nature, in some cases chaotic \citep[cf.][]{2024StrassmeieretalNC}.

\acknowledgements
This research was funded by the Hungarian National Research, Development, and Innovation Office grant KKP-143986. The financial support of the Austrian–Hungarian Action Foundation grant 117öu4 is also acknowledged.

\bibliography{Kovari_Kopal2024}

\begin{thebibliography}{63}
\expandafter\ifx\csname natexlab\endcsname\relax\def\natexlab#1{#1}\fi

\bibitem[{{Baliunas} {et~al.}(1996){Baliunas}, {Nesme-Ribes}, {Sokoloff}, \&
  {Soon}}]{1996ApJ...460..848B}
{Baliunas}, S.~L., {Nesme-Ribes}, E., {Sokoloff}, D., \& {Soon}, W.~H., {A
  Dynamo Interpretation of Stellar Activity Cycles}. 1996, {\it \apj}, {\bf
  460}, 848, DOI: 10.1086/177014

\bibitem[{{Berdyugina} \& {Tuominen}(1998)}]{1998A&A...336L..25B}
{Berdyugina}, S.~V. \& {Tuominen}, I., {Permanent active longitudes and
  activity cycles on RS CVn stars}. 1998, {\it \aap}, {\bf 336}, L25

\bibitem[{{Bois} {et~al.}(1991){Bois}, {Lanning}, \&
  {Mochnacki}}]{1991AJ....102.2079B}
{Bois}, B., {Lanning}, H.~H., \& {Mochnacki}, S.~W., {Spectroscopy of V471 Tau.
  II. Observations of the H(alpha) Line}. 1991, {\it \aj}, {\bf 102}, 2079,
  DOI: 10.1086/116032

\bibitem[{{Brun} \& {Browning}(2017)}]{2017LRSP...14....4B}
{Brun}, A.~S. \& {Browning}, M.~K., {Magnetism, dynamo action and the
  solar-stellar connection}. 2017, {\it Living Reviews in Solar Physics}, {\bf
  14}, 4, DOI: 10.1007/s41116-017-0007-8

\bibitem[{{Buccino} \& {Mauas}(2009)}]{2009A&A...495..287B}
{Buccino}, A.~P. \& {Mauas}, P.~J.~D., {Long-term chromospheric activity of
  non-eclipsing RS CVn-type stars}. 2009, {\it \aap}, {\bf 495}, 287, DOI:
  10.1051/0004-6361:200810985

\bibitem[{{Bushby}(2003)}]{2003MNRAS.338..655B}
{Bushby}, P.~J., {Strong asymmetry in stellar dynamos}. 2003, {\it \mnras},
  {\bf 338}, 655, DOI: 10.1046/j.1365-8711.2003.06080.x

\bibitem[{{Carroll} {et~al.}(2008){Carroll}, {Kopf}, \&
  {Strassmeier}}]{2008A&A...488..781C}
{Carroll}, T.~A., {Kopf}, M., \& {Strassmeier}, K.~G., {A fast method for
  Stokes profile synthesis. Radiative transfer modeling for ZDI and Stokes
  profile inversion}. 2008, {\it \aap}, {\bf 488}, 781, DOI:
  10.1051/0004-6361:200809981

\bibitem[{{Carroll} {et~al.}(2012){Carroll}, {Strassmeier}, {Rice}, \&
  {K{\"u}nstler}}]{2012A&A...548A..95C}
{Carroll}, T.~A., {Strassmeier}, K.~G., {Rice}, J.~B., \& {K{\"u}nstler}, A.,
  {The magnetic field topology of the weak-lined T Tauri star V410 Tauri. New
  strategies for Zeeman-Doppler imaging}. 2012, {\it \aap}, {\bf 548}, A95,
  DOI: 10.1051/0004-6361/201220215

\bibitem[{{Charbonneau} {et~al.}(2001){Charbonneau}, {McIntosh}, {Liu}, \&
  {Bogdan}}]{2001SoPh..203..321C}
{Charbonneau}, P., {McIntosh}, S.~W., {Liu}, H.-L., \& {Bogdan}, T.~J.,
  {Avalanche models for solar flares (Invited Review)}. 2001, {\it \solphys},
  {\bf 203}, 321, DOI: 10.1023/A:1013301521745

\bibitem[{{Cherkis} \& {Lyutikov}(2021)}]{2021ApJ...923...13C}
{Cherkis}, S.~A. \& {Lyutikov}, M., {Magnetic Topology in Coupled Binaries,
  Spin-orbital Resonances, and Flares}. 2021, {\it \apj}, {\bf 923}, 13, DOI:
  10.3847/1538-4357/ac29b8

\bibitem[{{Collier Cameron} \& {Robinson}(1989)}]{1989MNRAS.238..657C}
{Collier Cameron}, A. \& {Robinson}, R.~D., {Fast H{\ensuremath{\alpha}}
  variations on a rapidly rotating, cool main-sequence star-II. Cloud formation
  and ejection.} 1989, {\it \mnras}, {\bf 238}, 657, DOI:
  10.1093/mnras/238.2.657

\bibitem[{{Decampli} \& {Baliunas}(1979)}]{1979ApJ...230..815D}
{Decampli}, W.~M. \& {Baliunas}, S.~L., {What tides and flares do to RS Canum
  Venaticorum binaries.} 1979, {\it \apj}, {\bf 230}, 815, DOI: 10.1086/157140

\bibitem[{{Donati} \& {Collier Cameron}(1997)}]{1997MNRAS.291....1D}
{Donati}, J.~F. \& {Collier Cameron}, A., {Differential rotation and magnetic
  polarity patterns on AB Doradus}. 1997, {\it \mnras}, {\bf 291}, 1, DOI:
  10.1093/mnras/291.1.1

\bibitem[{{Ferreira} \& {Mendoza-Brice{\~n}o}(2005)}]{2005A&A...433.1055F}
{Ferreira}, J.~M. \& {Mendoza-Brice{\~n}o}, C.~A., {Coronal mass transfer in
  interbinary loops}. 2005, {\it \aap}, {\bf 433}, 1055, DOI:
  10.1051/0004-6361:20041926

\bibitem[{{Forbes}(2010)}]{2010hssr.book..159F}
{Forbes}, T., {Models of coronal mass ejections and flares}. 2010, in {\it
  Heliophysics: Space Storms and Radiation: Causes and Effects}, ed. C.~J.
  {Schrijver} \& G.~L. {Siscoe}, 159

\bibitem[{{Guinan} \& {Gim{\'e}nez}(1993)}]{1993ASSL..177...51G}
{Guinan}, E.~F. \& {Gim{\'e}nez}, A., {Magnetic Activity in Close Binaries}.
  1993, in Astrophysics and Space Science Library, Vol. {\bf  177}, {\it
  Astrophysics and Space Science Library}, ed. J.~{Sahade}, G.~E. {McCluskey},
  \& Y.~{Kondo}, 51

\bibitem[{{Guinan} {et~al.}(1986){Guinan}, {Wacker}, {Baliunas}, {Loesser}, \&
  {Raymond}}]{1986ESASP.263..197G}
{Guinan}, E.~F., {Wacker}, S.~W., {Baliunas}, S.~L., {Loesser}, J.~G., \&
  {Raymond}, J.~C., {Evidence of large-scale structures in the atmosphere of
  the active K-dwarf component of V471 Tauri.} 1986, in ESA Special
  Publication, Vol. {\bf  263}, {\it New Insights in Astrophysics. Eight Years
  of UV Astronomy with IUE}, ed. E.~J. {Rolfe} \& R.~{Wilson}, 197--200

\bibitem[{{Holzwarth} \& {Sch{\"u}ssler}(2000)}]{2000AN....321..175H}
{Holzwarth}, V. \& {Sch{\"u}ssler}, M., {Stability of magnetic flux tubes in
  close binary stars}. 2000, {\it Astronomische Nachrichten}, {\bf 321}, 175,
  DOI: 10.1002/1521-3994(200008)321:3<175::AID-ASNA175>3.0.CO;2-V

\bibitem[{{Holzwarth} \&
  {Sch{\"u}ssler}(2003{\natexlab{a}})}]{2003A&A...405..291H}
{Holzwarth}, V. \& {Sch{\"u}ssler}, M., {Dynamics of magnetic flux tubes in
  close binary stars. I. Equilibrium and stability properties}.
  2003{\natexlab{a}}, {\it \aap}, {\bf 405}, 291, DOI:
  10.1051/0004-6361:20030582

\bibitem[{{Holzwarth} \&
  {Sch{\"u}ssler}(2003{\natexlab{b}})}]{2003A&A...405..303H}
{Holzwarth}, V. \& {Sch{\"u}ssler}, M., {Dynamics of magnetic flux tubes in
  close binary stars. II. Nonlinear evolution and surface distributions}.
  2003{\natexlab{b}}, {\it \aap}, {\bf 405}, 303, DOI:
  10.1051/0004-6361:20030584

\bibitem[{{Hood} \& {Priest}(1979)}]{1979SoPh...64..303H}
{Hood}, A.~W. \& {Priest}, E.~R., {Kink Instability of Solar Coronal Loops as
  the Cause of Solar Flares}. 1979, {\it \solphys}, {\bf 64}, 303, DOI:
  10.1007/BF00151441

\bibitem[{{Hummel} \& {Beasley}(2020)}]{2020IAUS..354..418H}
{Hummel}, C.~A. \& {Beasley}, A., {Linking radio flares with spots on the
  active binary UX Arietis}. 2020, in IAU Symposium, Vol. {\bf  354}, {\it
  Solar and Stellar Magnetic Fields: Origins and Manifestations}, ed.
  A.~{Kosovichev}, S.~{Strassmeier}, \& M.~{Jardine}, 418--420

\bibitem[{{K{\"a}pyl{\"a}} {et~al.}(2023){K{\"a}pyl{\"a}}, {Browning}, {Brun},
  {Guerrero}, \& {Warnecke}}]{2023SSRv..219...58K}
{K{\"a}pyl{\"a}}, P.~J., {Browning}, M.~K., {Brun}, A.~S., {Guerrero}, G., \&
  {Warnecke}, J., {Simulations of Solar and Stellar Dynamos and Their
  Theoretical Interpretation}. 2023, {\it \ssr}, {\bf 219}, 58, DOI:
  10.1007/s11214-023-01005-6

\bibitem[{{K\H{o}v\'ari} \& {Bartus}(1997)}]{1997A&A...323..801K}
{K\H{o}v\'ari}, {\mbox Zs}. \& {Bartus}, J., {Testing the stability and
  reliability of starspot modelling.} 1997, {\it \aap}, {\bf 323}, 801

\bibitem[{{K{\H{o}}v{\'a}ri} {et~al.}(2012){K{\H{o}}v{\'a}ri}, {Korhonen},
  {Kriskovics}, {Vida}, {Donati}, {Le Coroller}, {Monnier}, {Pedretti}, \&
  {Petit}}]{2012A&A...539A..50K}
{K{\H{o}}v{\'a}ri}, {\mbox Zs}., {Korhonen}, H., {Kriskovics}, L., {et~al.},
  {Measuring differential rotation of the K-giant {\ensuremath{\zeta}}
  Andromedae}. 2012, {\it \aap}, {\bf 539}, A50, DOI:
  10.1051/0004-6361/201118177

\bibitem[{{K{\H{o}}v{\'a}ri} {et~al.}(2015){K{\H{o}}v{\'a}ri}, {Kriskovics},
  {K{\"u}nstler}, {Carroll}, {Strassmeier}, {Vida}, {Ol{\'a}h}, {Bartus}, \&
  {Weber}}]{2015A&A...573A..98K}
{K{\H{o}}v{\'a}ri}, {\mbox Zs}., {Kriskovics}, L., {K{\"u}nstler}, A.,
  {et~al.}, {Antisolar differential rotation of the K1-giant
  {\ensuremath{\sigma}} Geminorum revisited}. 2015, {\it \aap}, {\bf 573}, A98,
  DOI: 10.1051/0004-6361/201424138

\bibitem[{{K{\H{o}}v{\'a}ri} {et~al.}(2021){K{\H{o}}v{\'a}ri}, {Kriskovics},
  {Ol{\'a}h}, {Odert}, {Leitzinger}, {Seli}, {Vida}, {Borkovits}, \&
  {Carroll}}]{2021A&A...650A.158K}
{K{\H{o}}v{\'a}ri}, {\mbox Zs}., {Kriskovics}, L., {Ol{\'a}h}, K., {et~al.}, {A
  confined dynamo: Magnetic activity of the K-dwarf component in the
  pre-cataclysmic binary system V471 Tauri}. 2021, {\it \aap}, {\bf 650}, A158,
  DOI: 10.1051/0004-6361/202140707

\bibitem[{{K{\H{o}}v{\'a}ri} {et~al.}(2017){K{\H{o}}v{\'a}ri}, {Ol{\'a}h},
  {Kriskovics}, {Vida}, {Forg{\'a}cs-Dajka}, \&
  {Strassmeier}}]{2017AN....338..903K}
{K{\H{o}}v{\'a}ri}, {\mbox Zs}., {Ol{\'a}h}, K., {Kriskovics}, L., {et~al.},
  {Rotation-differential rotation relationships for late-type single and binary
  stars from Doppler imaging}. 2017, {\it Astronomische Nachrichten}, {\bf
  338}, 903, DOI: 10.1002/asna.201713400

\bibitem[{{K{\H{o}}v{\'a}ri} {et~al.}(2004){K{\H{o}}v{\'a}ri}, {Strassmeier},
  {Granzer}, {Weber}, {Ol{\'a}h}, \& {Rice}}]{2004A&A...417.1047K}
{K{\H{o}}v{\'a}ri}, {\mbox Zs}., {Strassmeier}, K.~G., {Granzer}, T., {et~al.},
  {Doppler imaging of stellar surface structure. XXII. Time-series mapping of
  the young rapid rotator LQ Hydrae}. 2004, {\it \aap}, {\bf 417}, 1047, DOI:
  10.1051/0004-6361:20034187

\bibitem[{{Kitchatinov} \& {R{\"u}diger}(2004)}]{2004AN....325..496K}
{Kitchatinov}, L.~L. \& {R{\"u}diger}, G., {Anti-solar differential rotation}.
  2004, {\it Astronomische Nachrichten}, {\bf 325}, 496, DOI:
  10.1002/asna.200410297

\bibitem[{{Kopp} \& {Pneuman}(1976)}]{1976SoPh...50...85K}
{Kopp}, R.~A. \& {Pneuman}, G.~W., {Magnetic reconnection in the corona and the
  loop prominence phenomenon.} 1976, {\it \solphys}, {\bf 50}, 85, DOI:
  10.1007/BF00206193

\bibitem[{{Kouzuma}(2019)}]{2019PASJ...71...21K}
{Kouzuma}, S., {Starspots in contact and semi-detached binary systems}. 2019,
  {\it \pasj}, {\bf 71}, 21, DOI: 10.1093/pasj/psy140

\bibitem[{{Kundra} {et~al.}(2022){Kundra}, {Hamb{\'a}lek}, {Vanaverbeke},
  {Dubovsk{\'y}}, {Logie}, {Rau}, \& {Dubois}}]{2022MNRAS.517.5358K}
{Kundra}, E., {Hamb{\'a}lek}, {\v{L}}., {Vanaverbeke}, S., {et~al.},
  {Variability of eclipse timing: the case of V471 Tauri}. 2022, {\it \mnras},
  {\bf 517}, 5358, DOI: 10.1093/mnras/stac2812

\bibitem[{{Lestrade} {et~al.}(1985){Lestrade}, {Mutel}, {Preston}, \&
  {Phillips}}]{1985ASSL..116..275L}
{Lestrade}, J.~F., {Mutel}, R.~L., {Preston}, R.~A., \& {Phillips}, R.~B.,
  {High-angular Resolution Observations of Stellar Binary Systems}. 1985, in
  Astrophysics and Space Science Library, Vol. {\bf  116}, {\it Radio Stars},
  ed. R.~M. {Hjellming} \& D.~M. {Gibson}, 275--282

\bibitem[{{Luger} {et~al.}(2021){Luger}, {Foreman-Mackey}, {Hedges}, \&
  {Hogg}}]{2021AJ....162..123L}
{Luger}, R., {Foreman-Mackey}, D., {Hedges}, C., \& {Hogg}, D.~W., {Mapping
  Stellar Surfaces. I. Degeneracies in the Rotational Light-curve Problem}.
  2021, {\it \aj}, {\bf 162}, 123, DOI: 10.3847/1538-3881/abfdb8

\bibitem[{{Mart{\'\i}nez} {et~al.}(2022){Mart{\'\i}nez}, {Mauas}, \&
  {Buccino}}]{2022MNRAS.512.4835M}
{Mart{\'\i}nez}, C.~I., {Mauas}, P.~J.~D., \& {Buccino}, A.~P., {Activity
  cycles in RS CVn-type stars}. 2022, {\it \mnras}, {\bf 512}, 4835, DOI:
  10.1093/mnras/stac755

\bibitem[{{Moss} {et~al.}(1995){Moss}, {Barker}, {Brandenburg}, \&
  {Tuominen}}]{1995A&A...294..155M}
{Moss}, D., {Barker}, D.~M., {Brandenburg}, A., \& {Tuominen}, I.,
  {Nonaxisymmetric dynamo solutions and extended starspots on late-type stars.}
  1995, {\it \aap}, {\bf 294}, 155

\bibitem[{{Moss} {et~al.}(2002){Moss}, {Piskunov}, \&
  {Sokoloff}}]{2002A&A...396..885M}
{Moss}, D., {Piskunov}, N., \& {Sokoloff}, D., {Nonaxisymmetric cool spot
  distributions and dynamo action in close binaries}. 2002, {\it \aap}, {\bf
  396}, 885, DOI: 10.1051/0004-6361:20021370

\bibitem[{{Moss} \& {Tuominen}(1997)}]{1997A&A...321..151M}
{Moss}, D. \& {Tuominen}, I., {Magnetic field generation in close binary
  systems.} 1997, {\it \aap}, {\bf 321}, 151

\bibitem[{{Ol{\'a}h}(2006)}]{2006Ap&SS.304..145O}
{Ol{\'a}h}, K., {Active Longitudes in Close Binaries}. 2006, {\it \apss}, {\bf
  304}, 145, DOI: 10.1007/s10509-006-9096-x

\bibitem[{{Ol{\'a}h} {et~al.}(2021){Ol{\'a}h}, {K{\H{o}}v{\'a}ri},
  {G{\"u}nther}, {Vida}, {Gaulme}, {Seli}, \& {P{\'a}l}}]{2021A&A...647A..62O}
{Ol{\'a}h}, K., {K{\H{o}}v{\'a}ri}, {\mbox Zs}., {G{\"u}nther}, M.~N.,
  {et~al.}, {Toward the true number of flaring giant stars in the Kepler field.
  Are their flaring specialities associated with their being giant stars?}
  2021, {\it \aap}, {\bf 647}, A62, DOI: 10.1051/0004-6361/202039674

\bibitem[{{Ol{\'a}h} {et~al.}(2009){Ol{\'a}h}, {Koll{\'a}th}, {Granzer},
  {Strassmeier}, {Lanza}, {J{\"a}rvinen}, {Korhonen}, {Baliunas}, {Soon},
  {Messina}, \& {Cutispoto}}]{2009A&A...501..703O}
{Ol{\'a}h}, K., {Koll{\'a}th}, Z., {Granzer}, T., {et~al.}, {Multiple and
  changing cycles of active stars. II. Results}. 2009, {\it \aap}, {\bf 501},
  703, DOI: 10.1051/0004-6361/200811304

\bibitem[{{Ol{\'a}h} \& {K{\"o}v{\'a}ri}(1997)}]{1997eaun.book..151O}
{Ol{\'a}h}, K. \& {K{\"o}v{\'a}ri}, {\mbox Zs}., {Doubly Active Binaries}.
  1997, in {\it The Earth and the Universe}, ed. G.~{Asteriadis},
  A.~{Bantelas}, M.~E. {Contadakis}, K.~{Katsambalos}, A.~{Papdimitriou}, \&
  I.~N. {Tziavos}, 151

\bibitem[{{Ol{\'a}h} {et~al.}(2014){Ol{\'a}h}, {Mo{\'o}r}, {K{\H{o}}v{\'a}ri},
  {Granzer}, {Strassmeier}, {Kriskovics}, \& {Vida}}]{2014A&A...572A..94O}
{Ol{\'a}h}, K., {Mo{\'o}r}, A., {K{\H{o}}v{\'a}ri}, {\mbox Zs}., {et~al.},
  {Magnitude-range brightness variations of overactive K giants}. 2014, {\it
  \aap}, {\bf 572}, A94, DOI: 10.1051/0004-6361/201424695

\bibitem[{{Ol{\'a}h} {et~al.}(2013){Ol{\'a}h}, {Mo{\'o}r}, {Strassmeier},
  {Borkovits}, \& {Granzer}}]{2013AN....334..625O}
{Ol{\'a}h}, K., {Mo{\'o}r}, A., {Strassmeier}, K.~G., {Borkovits}, T., \&
  {Granzer}, T., {Long-term photometry of three active red giants in close
  binary systems: V2253 Oph, IT Com and IS Vir}. 2013, {\it Astronomische
  Nachrichten}, {\bf 334}, 625, DOI: 10.1002/asna.201211846

\bibitem[{{{\"O}zdarcan}(2021)}]{2021PASA...38...27O}
{{\"O}zdarcan}, O., {Spectroscopic and photometric analysis of 21
  chromospherically active variables: Activity cycles and differential
  rotation}. 2021, {\it \pasa}, {\bf 38}, e027, DOI: 10.1017/pasa.2021.21

\bibitem[{{Peterson} {et~al.}(2010){Peterson}, {Mutel}, {G{\"u}del}, \&
  {Goss}}]{2010Natur.463..207P}
{Peterson}, W.~M., {Mutel}, R.~L., {G{\"u}del}, M., \& {Goss}, W.~M., {A large
  coronal loop in the Algol system}. 2010, {\it \nat}, {\bf 463}, 207, DOI:
  10.1038/nature08643

\bibitem[{{Priest}(2014)}]{2014masu.book.....P}
{Priest}, E. 2014, {\it {Magnetohydrodynamics of the Sun}}

\bibitem[{{Roettenbacher} {et~al.}(2016){Roettenbacher}, {Monnier}, {Korhonen},
  {Aarnio}, {Baron}, {Che}, {Harmon}, {K{\H{o}}v{\'a}ri}, {Kraus}, {Schaefer},
  {Torres}, {Zhao}, {Ten Brummelaar}, {Sturmann}, \&
  {Sturmann}}]{2016Natur.533..217R}
{Roettenbacher}, R.~M., {Monnier}, J.~D., {Korhonen}, H., {et~al.}, {No
  Sun-like dynamo on the active star {\ensuremath{\zeta}} Andromedae from
  starspot asymmetry}. 2016, {\it \nat}, {\bf 533}, 217, DOI:
  10.1038/nature17444

\bibitem[{{Roettenbacher} {et~al.}(2017){Roettenbacher}, {Monnier}, {Korhonen},
  {Harmon}, {Baron}, {Hackman}, {Henry}, {Schaefer}, {Strassmeier}, {Weber}, \&
  {ten Brummelaar}}]{2017ApJ...849..120R}
{Roettenbacher}, R.~M., {Monnier}, J.~D., {Korhonen}, H., {et~al.},
  {Contemporaneous Imaging Comparisons of the Spotted Giant
  {\ensuremath{\sigma}} Geminorum Using Interferometric, Spectroscopic, and
  Photometric Data}. 2017, {\it \apj}, {\bf 849}, 120, DOI:
  10.3847/1538-4357/aa8ef7

\bibitem[{{Salter} {et~al.}(2010){Salter}, {K{\'o}sp{\'a}l}, {Getman},
  {Hogerheijde}, {van Kempen}, {Carpenter}, {Blake}, \&
  {Wilner}}]{2010A&A...521A..32S}
{Salter}, D.~M., {K{\'o}sp{\'a}l}, {\'A}., {Getman}, K.~V., {et~al.},
  {Recurring millimeter flares as evidence for star-star magnetic reconnection
  events in the DQ Tauri PMS binary system}. 2010, {\it \aap}, {\bf 521}, A32,
  DOI: 10.1051/0004-6361/201015197

\bibitem[{{Samujllo} \& {Kiraga}(2007)}]{2007AcA....57..347S}
{Samujllo}, M. \& {Kiraga}, M., {Evolution of Magnetic Flux Tubes in Convective
  Envelopes of Close Binary Stars}. 2007, {\it \actaa}, {\bf 57}, 347

\bibitem[{{Savanov}(2013)}]{2013IAUS..294..257S}
{Savanov}, I.~S., {Starspot detection and properties}. 2013, in IAU Symposium,
  Vol. {\bf  294}, {\it Solar and Astrophysical Dynamos and Magnetic Activity},
  ed. A.~G. {Kosovichev}, E.~{de Gouveia Dal Pino}, \& Y.~{Yan}, 257--268

\bibitem[{{Schrijver} \& {Zwaan}(1991)}]{1991A&A...251..183S}
{Schrijver}, C.~J. \& {Zwaan}, C., {Activity in tidally interacting binaries.}
  1991, {\it \aap}, {\bf 251}, 183

\bibitem[{{Sethi} \& {Martin}(2024)}]{2024MNRAS.529.4442S}
{Sethi}, R. \& {Martin}, D.~V., {Tight stellar binaries favour active
  longitudes at sub- and antistellar points}. 2024, {\it \mnras}, {\bf 529},
  4442, DOI: 10.1093/mnras/stae717

\bibitem[{{Siarkowski} {et~al.}(1996){Siarkowski}, {Pres}, {Drake}, {White}, \&
  {Singh}}]{1996ApJ...473..470S}
{Siarkowski}, M., {Pres}, P., {Drake}, S.~A., {White}, N.~E., \& {Singh},
  K.~P., {Corona(e) of AR Lacertae. II. The Spatial Structure}. 1996, {\it
  \apj}, {\bf 473}, 470, DOI: 10.1086/178159

\bibitem[{{Simon} {et~al.}(1980){Simon}, {Linsky}, \&
  {Schiffer}}]{1980ApJ...239..911S}
{Simon}, T., {Linsky}, J.~L., \& {Schiffer}, F.~H., I., {IUE spectra of a flare
  in the RS Canum Venaticorum-type system UX Arietis.} 1980, {\it \apj}, {\bf
  239}, 911, DOI: 10.1086/158178

\bibitem[{{Singh} \& {Pandey}(2022)}]{2022ApJ...934...20S}
{Singh}, G. \& {Pandey}, J.~C., {An X-Ray Study of Coronally Connected Active
  Eclipsing Binaries}. 2022, {\it \apj}, {\bf 934}, 20, DOI:
  10.3847/1538-4357/ac7716

\bibitem[{{Sokoloff} \& {Piskunov}(2002)}]{2002MNRAS.334..925S}
{Sokoloff}, D. \& {Piskunov}, N., {Swing excitation and magnetic activity in
  close binary systems}. 2002, {\it \mnras}, {\bf 334}, 925, DOI:
  10.1046/j.1365-8711.2002.05583.x

\bibitem[{{Strassmeier} {et~al.}(2024){Strassmeier}, {K{\H{o}}v{\'a}ri},
  {Weber}, \& {Granzer}}]{2024StrassmeieretalNC}
{Strassmeier}, K.~G., {K{\H{o}}v{\'a}ri}, {\mbox Zs}., {Weber}, M., \&
  {Granzer}, T., {Long-term Doppler imaging of the star XX Trianguli indicates
  chaotic nonperiodic dynamo}. 2024, {\it Nature Communications}, submitted,
  DOI: 10.21203/rs.3.rs-3118867/v1

\bibitem[{{van den Oord}(1988)}]{1988A&A...205..167V}
{van den Oord}, G.~H.~J., {Filament support and flares in binaries}. 1988, {\it
  \aap}, {\bf 205}, 167

\bibitem[{{Young} {et~al.}(1991){Young}, {Rottler}, \&
  {Skumanich}}]{1991ApJ...378L..25Y}
{Young}, A., {Rottler}, L., \& {Skumanich}, A., {Evidence for External Plasma
  around the K Dwarf Component of the Eclipsing Binary V471 Tauri}. 1991, {\it
  \apjl}, {\bf 378}, L25, DOI: 10.1086/186133

\bibitem[{{Young} {et~al.}(1988){Young}, {Skumanich}, \&
  {Paylor}}]{1988ApJ...334..397Y}
{Young}, A., {Skumanich}, A., \& {Paylor}, V., {Fluorescence-induced
  Chromospheric H alpha Emission from the K Dwarf Component of V471 Tauri. I.
  The 1983 Epoch}. 1988, {\it \apj}, {\bf 334}, 397, DOI: 10.1086/166844

\end{thebibliography}

\end{document}